\documentclass[prl,aps,twocolumn,amsfonts,showpacs,superscriptaddress,floatfix]{revtex4}

\usepackage{epsfig}
\usepackage{psfrag}
\usepackage{color}
\usepackage{graphicx}
\usepackage{subfigure}

\begin{document}


\title{Kondo Physics and Exact Solvability of Double Dots Systems}
\author{Robert M. Konik}
\affiliation{Condensed Matter Physics and Material Science Department, Brookhaven National Laboratory, Upton,
NY 11973}

\date{\today}

\begin{abstract}
We study two double dot systems, one with dots in parallel and one with dots in series,
and argue they admit an exact solution via the Bethe ansatz.  In the case of parallel dots
we exploit the exact solution
to extract the behavior of the linear response conductance.  The linear response conductance
of the parallel dot system possesses multiple Kondo effects, including
a Kondo effect enhanced by a nonpertubative antiferromagnetic RKKY interaction, 
has conductance zeros in the mixed valence regime, and obeys a non-trivial form of the Friedel sum rule.
\end{abstract}
\maketitle

\newcommand{\del}{\partial}
\newcommand{\ep}{\epsilon}
\newcommand{\clsd}{c_{l\sig}^\dagger}
\newcommand{\cls}{c_{l\sig}}
\newcommand{\cesd}{c_{e\sig}^\dagger}
\newcommand{\ces}{c_{e\sig}}
\newcommand{\up}{\uparrow}
\newcommand{\down}{\downarrow}
\newcommand{\il}{\int^{\tilde{Q}}_Q d\la~}
\newcommand{\ilp}{\int^{\tilde{Q}}_Q d\la '}
\newcommand{\ik}{\int^{B}_{-D} dk~}
\newcommand{\ila}{\int d\la~}
\newcommand{\ilpa}{\int d\la '}
\newcommand{\ika}{\int dk~}
\newcommand{\tQ}{\tilde{Q}}
\newcommand{\rh}{\rho_{\rm bulk}}
\newcommand{\ri}{\rho^{\rm imp}}
\newcommand{\sh}{\sig_{\rm bulk}}
\newcommand{\si}{\sig^{\rm imp}}
\newcommand{\rph}{\rho_{p/h}}
\newcommand{\sph}{\sig_{p/h}}
\newcommand{\rp}{\rho_{p}}
\newcommand{\sip}{\sig_{p}}
\newcommand{\drph}{\delta\rho_{p/h}}
\newcommand{\dsph}{\delta\sig_{p/h}}
\newcommand{\drp}{\delta\rho_{p}}
\newcommand{\dsp}{\delta\sig_{p}}
\newcommand{\drh}{\delta\rho_{h}}
\newcommand{\dsh}{\delta\sig_{h}}
\newcommand{\enp}{\ep^+}
\newcommand{\enm}{\ep^-}
\newcommand{\enpm}{\ep^\pm}
\newcommand{\enph}{\ep^+_{\rm bulk}}
\newcommand{\enmh}{\ep^-_{\rm bulk}}
\newcommand{\enpi}{\ep^+_{\rm imp}}
\newcommand{\enmi}{\ep^-_{\rm imp}}
\newcommand{\enh}{\ep_{\rm bulk}}
\newcommand{\eni}{\ep_{\rm imp}}
\newcommand{\sig}{\sigma}
\newcommand{\la}{\lambda}
\newcommand{\ua}{\uparrow}
\newcommand{\da}{\downarrow}
\newcommand{\ed}{\epsilon_d}

The continual advance in the ability to engineer devices on the nanoscale level
has led to the recent fabrication of double quantum dot (DQD) devices 
\cite{RKKY1,RKKY2,scdots1,scdots2,scdots3,scdots4,cndots1,cndots2}, both from
semiconducting heterostructures \cite{RKKY1,RKKY2,scdots1,scdots2,scdots3,scdots4}
and from carbon nanotubes \cite{cndots1,cndots2}.
Like their single dot cousins \cite{gold,kondo}, these devices are highly tunable.  
This tunability makes DQDs both a leading candidate for a solid state realization of a quantum qubit
and an ideal laboratory to observe strongly correlated Kondo phenomena.

Single dot devices have long been known to exhibit Kondo physics \cite{gold,kondo}.  
By the use of a gate voltage, the number of electrons on
the dot can be adjusted so that it is odd.  By virtue of Kramers degeneracy, the dot, 
through hybridization with electrons
in connecting leads, becomes
a realization of a single impurity Kondo system.  Transport measurements are the most striking
signatures of Kondo physics in such devices, where, for example, the observed finite temperature linear 
response conductance \cite{gold} matches that predicted theoretically \cite{costi,konik}.

With different possible dot geometries, 
Kondo physics in DQD devices is necessarily more rich.
Dots arranged in series have been shown
to exhibit Kondo physics in competition with 
an effective 
Ruderman-Kittel-Kasuya-Yosida (RKKY) coupling between dots \cite{RKKY1,RKKY2,glazman,oreg,lopez}. 
And  
in parallel dot systems, Kondo physics has been observed \cite{scdots3,scdots4} where
it is expected to appear in conjunction
with interference effects \cite{lopez,meir,kawa}.

In this letter, we present a powerful theoretical approach to DQD systems. 
We study two models of such systems, one with dots in parallel, one with dots in series.
We argue both models admit exact solutions via the Bethe ansatz.  
The model of dots in series possesses an SU(4) symmetry \cite{su(4)} and is relevant 
to the study of the Kondo effect in carbon nanotubes \cite{tubes}.
For the model of dots in parallel, we explicitly demonstrate 
that transport properties may be extracted from the exact solution.  
Transport in the system of parallel dots exhibits two interesting features:
i) at the particle-hole symmetric point, we find a {\it cooperative} 
combination of Kondo and RKKY physics where a non-perturbative antiferromagnetic RKKY effect
serves to mediate the formation of a Kondo like singlet; and ii) in the dots' mixed valence regime,
we discover a
form of the Friedel sum rule where contributions to the scattering phase come both from 
electrons in the dots {\it and the leads}.

\noindent {\bf Models Defined:} We examine 
two generalized Anderson models coupling two leads ($l=1,2$) to 
two DQDs ($\alpha =1,2$) in two distinct configurations.  In the first configuration (denoted by
{\bf PD}), the dots
are arranged in parallel.  The corresponding Hamiltonian is taken to be
${\cal H}_1 = {\cal H}_0+ {\cal H}_{\rm1 int}$
with 
${\cal H}_0 = -i\sum_{l\sigma}\int^\infty_{-\infty} dx c^\dagger_{l\sigma}\partial_xc_{l\sigma}
+ \sum_{\sigma\alpha}\epsilon_{d\alpha}n_{\sigma\alpha}$
and 
\begin{equation}\label{ei}
{\cal H}_{\rm 1 int} = \sum_{l\sigma\alpha}
V_{l\alpha}(c^\dagger_{\sigma l}d_{\sigma\alpha} + {\rm h.c.}) 
+ \sum_{\alpha\alpha'} U_{\alpha\alpha'} n_{\ua\alpha}n_{\da\alpha}.
\end{equation}
Here the $c_{l\sigma}/d_{\alpha\sigma}$ specify electrons living in the leads and the dots.
$V_{l\alpha}$ measures
the tunneling strength between the dot $\alpha$ and lead $l$.  $U_{\alpha\alpha'}$ characterizes
the Coulombic repulsion between electrons of opposite spin living on dots $\alpha$ and $\alpha'$.

\begin{figure}[tbh]
\centerline{\psfig{figure=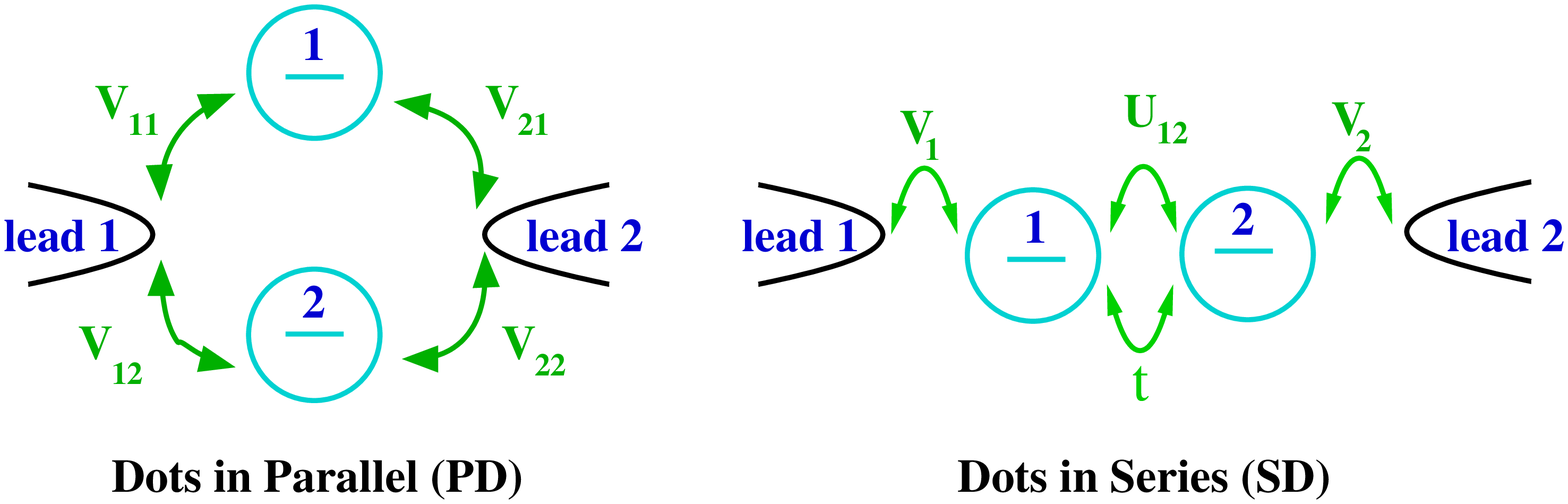,height=1.in,width=3in}}
\caption{A schematic of two DQD systems.}
\end{figure}

In the second model (denoted {\bf SD}) the dots are arranged in series.  The corresponding Hamiltonian has
a similar form to the first ${\cal H}_2 = P({\cal H}_0+ {\cal H}_{\rm 2 int})P$.  Here
${\cal H}_{\rm 2 int}$ is of the form 
\begin{eqnarray}\label{eii}
{\cal H}_{\rm 2 int} &=& \sum_{\sigma,l}V_l(c^\dagger_{l\sigma}d_{\sigma l} + {\rm h.c.}) 
+ \!\!\!\!\sum_{(\alpha\sigma)\neq(\alpha'\sigma')}\!\!\!\!
n_{\alpha\sigma}n_{\alpha'\sigma'}\cr
&& \hskip -.2in \times\bigg(U + i 
\lim_{a \rightarrow 0}\sum_{l\sigma}\int^a_{-a}dx c^\dagger_{l\sigma}(x)\partial_x c_{l\sigma}(x)\bigg).
\end{eqnarray}
$P$ is a projection operator which forbids total occupancy of both dots from exceeding
two electrons.  In a regime where fewer than two electrons sit on the dots, the role
played by both the projection operator and the correlated hopping terms will be minimal.
We note that the addition of an interdot hopping term, i.e. $\delta{\cal H}_2 = t_\perp(d^\dagger_1d_2 + d^\dagger_2d_1)$,
does not spoil integrability \cite{konik2}.

\noindent {\bf Integrability of PD:}
To analyze {\bf PD} (two dots in parallel) we first map the problem to an Anderson model
involving a single effective lead.  To do so we need to assume the ratio
of left/right lead couplings are equal, i.e. $V_{L\alpha}/V_{R\alpha}=V_{L\alpha'}/V_{R\alpha'}$.
Writing $c_{e/o} = (V_{L/R\alpha}c_L \pm V_{R/L\alpha}c_R)/\sqrt{2\Gamma_\alpha}$, with 
$\Gamma_\alpha = (V_{L\alpha}^2+V_{R\alpha}^2)/2$, the Hamiltonian factorizes
into an even and an odd sector, ${\cal H}_1 = {\cal H}_e + {\cal H}_o$.
Only ${\cal H}_e$ couples to the dot and 
is given by ${\cal H}_e = {\cal H}_{e0}+{\cal H}_{e\rm int}$ 
with 
${\cal H}_{e0} = -i\sum_{l\sigma}\int^\infty_{-\infty} dx c^\dagger_{e\sigma}\partial_xc_{e\sigma}
+ \sum_{\sigma\alpha}\epsilon_{d\alpha}n_{\sigma\alpha}$
and
\begin{equation}\label{eiii}
{\cal H}_{e\rm int} = 
\sum_{\sigma\alpha}\sqrt{2\Gamma_{\alpha}}(c^\dagger_{e\sigma\alpha}d_{\sigma\alpha} + {\rm h.c.}) 
+ \sum_{\alpha\alpha'} U_{\alpha\alpha'} n_{\ua\alpha}n_{\da\alpha}.
\end{equation}
In contrast, ${\cal H}_o$ is trivial:
${\cal H}_o = {\cal H}_{e0}(e\rightarrow o,\epsilon_{d\alpha}=0)$.
The integrability of {\bf PD} is then equivalent to the exact solvability of ${\cal H}_e$.

To determine under what conditions ${\cal H}_e$ admits eigenfunctions of the Bethe ansatz form,
we begin by computing both the
one and two electron eigenstates.  The one particle wave function appears as
\begin{eqnarray}\label{eiv}
|\psi_\sigma\rangle = \bigg[ \int^\infty_{-\infty}dx \{ g_\sigma (x) c^\dagger_\sigma(x) \} 
+ e_{\alpha\sigma} d^\dagger_{\alpha\sigma}\bigg]|0\rangle .
\end{eqnarray}
Solving the Schr\"odinger equation, ${\cal H}_e |\psi\rangle = q|\psi\rangle$, we find
$g_\sigma (x) = \theta (x) e^{iqx + i\delta/2} + \theta (-x) e^{iqx - i\delta/2}$
with $\delta (q)$, the impurity scattering phase, to be
$\delta (q) = -2\tan^{-1}(\sum_\alpha (\Gamma_\alpha/(q - \epsilon_{d\alpha}))$.
To compute the effective scattering between electrons, we study the
two particle eigenfunction with spin projection, $S_z=0$:
\begin{eqnarray}\label{evi}
|\psi\rangle &=& \bigg[ \int^{\infty}_{-\infty} dx_1dx_2 g(x_1,x_2)c^\dagger_\uparrow (x_1) 
c^\dagger_\downarrow (x_2)
+ \sum_{\alpha}\int^\infty_{-\infty} dx  \cr\cr
&& \hskip -.4in 
\big[e_\alpha(x) (c^\dagger_\uparrow (x) d^\dagger_{\alpha\da} - c^\dagger_\downarrow (x)d^\dagger_{\ua\alpha})\big]
+ \sum_{\alpha\alpha'} f_{\alpha\alpha'}d^\dagger_{\ua\alpha} d^\dagger_{\da\alpha'}\bigg]|0\rangle .
\end{eqnarray}
Again solving the Schr\"odinger equation ${\cal H}_e|\psi\rangle = (q+p)|\psi\rangle$ gives $g(x_1,x_2)$ to be
$g(x_1,x_2) = g_q (x_1)g_p(x_2)\phi(x_{12})
+ (x_1\leftrightarrow x_2)$,
with $x_{12}=x_1-x_2$.
Here $g_{q/p}(x)$ are
one particle wavefunctions with
energies $q/p$.  $\phi (x)$ governs the scattering when two electrons are interchanged.  It takes the
form $\phi(x) = 1 + i\gamma (q,p){\rm sign} (x)$.  We find that
$\gamma (q,p)$ is consistently determined to be
\begin{equation}\label{eviii}
\gamma (q,p) = {1\over q - p} \sum_{\alpha'} 
{2\Gamma_{\alpha'}U_{\alpha\alpha'} \over \epsilon_{d\alpha} +
\epsilon_{d\alpha'}+U_{\alpha\alpha'} - q - p},
\end{equation}
if either
\begin{eqnarray}\label{eix}
U_{\alpha\alpha '} = \delta_{\alpha \alpha'}U_{\alpha};~U_{\alpha} \Gamma_\alpha = c, ~
U_{\alpha} + 2\epsilon_{d\alpha} = c';
\end{eqnarray}
where $c$ and $c'$ are $\alpha$-independent constants or
\begin{eqnarray}\label{ex}
U_{\alpha\alpha'} &=& U;~~\Gamma_\alpha = \Gamma_{\alpha'};~~\epsilon_\alpha = \epsilon_{\alpha'}.
\end{eqnarray}
The first set of conditions 
describes single level dots absent an interdot coupling while the second 
dots with degenerate levels but with a highly finely tuned
interdot interaction.  We will thus focus on the first.

Exact solvability is predicated on how $\gamma (q,p)$ determines the scattering matrix of the two electrons.
The scattering matrix has the general spin (SU(2)) invariant form,
$S^{a'b'}_{ab} = b(p,q) I^{a'b'}_{ab}
+ c(p,q) P^{a'b'}_{ab}$,
where $a,b=\ua,\da$ and ${\bf I/P}$ are the identity/permutation matrices.
The coefficients, $b(p,q)$ and $c(p,q)$, are determined by $\gamma (p,q)$ from
the relation,
$b(p,q)-c(p,q) = \phi(x>0)/ \phi(x<0) = (1+i\gamma (p,q))/(1-i\gamma(p,q))$,
together with $b(p,q)+c(p,q) = 1$ which arises from considering the eigenfunction in
$S_z = \pm 1$ sector where interactions are absent.

In order for the Hamiltonian, ${\cal H}_e$, to be integrable, a minimal condition is that
the above S-matrix satisfy the
Yang-Baxter relation.  The Yang-Baxter relation governs the scattering of three electrons
and it enforces the equivalency of different scattering orders.
It is well known in the case of an $SU(2)$ symmetry that the validity of the Yang-Baxter relations 
is equivalent to the condition, 
$b(p,q)/c(p,q) = i(g(p) - g(q)),$
where $g(p)$ is an arbitrary function \cite{original}.  In the case at hand, $g(p)$ is given by
$g(p) = (p - \epsilon_{d\alpha} - U_{\alpha}/2)^2/(2\Gamma_\alpha U_{\alpha})$.

Having determined under 
what conditions the dot-lead Hamiltonian is exactly solvable, we are now in a position to
construct 
$N$-particle eigenstates in a controlled fashion.  An eigenfunction with spin, $S_z = N-2M$, 
is characterized
by a sea of N electrons each carrying momenta $\{ q_i\}^N_{i=1}$ and so total energy $E=\sum_i q_i$.  
In a periodic system of length L, integrability allows us to write down in a compact form the 
$q_i$-quantization conditions (the Bethe ansatz equations):
\begin{eqnarray}\label{exiii}
e^{iq_jL+i\delta(q_j)} &=& \prod^M_{\alpha = 1}{g(q_j)-\lambda_\alpha+i/2 \over g(q_j)-\lambda_\alpha-i/2};\cr\cr
\prod^N_{j=1}{\lambda_\alpha-g(q_j)+i/2 \over \lambda_\alpha-g(q_j)-i/2} 
&=& -\prod^M_{\beta=1}{\lambda_\alpha-\lambda_\beta+i \over \lambda_\alpha-\lambda_\beta-i}.
\end{eqnarray}
These equations are identical to those for the ordinary Anderson model \cite{original} but for the form 
of $\delta (q)$.  The M $\lambda_\alpha$'s appearing in the above equations are indicative of the
spin degrees of freedom.  

There are two integrable generalizations of the {\bf PD} model: i) N-dots in
parallel and ii) a pair of dots in a T-junction.  These systems, to be integrable, must satisfy
a set of constraints similar to Eqns. 7 and 8.

\noindent {\bf Integrability of SD:}
We demonstrate the generic integrability of {\bf SD} along similar lines --
it was already known to be exactly solvable for infinite $U$ \cite{schlottmann}.
Here, however,
we do not transform to an even/odd sector.  We thus have four different types of fermions/dot degrees
of freedom.  This will ultimately lead to the model having an exact $SU(4)$ symmetry at all energy scales.

Constructing the two particle eigenfunctions as before leads to an $SU(4)$ S-matrix
provided $V_l = V$ for all $l$ and $\epsilon_{d\alpha}= \epsilon_{d}$ for all $\alpha$.
(This latter constraint can be relaxed enabling one to study non-degenerate dots where
the $SU(4)$ symmetry is broken, say by a magnetic field or a gate voltage.)
The S-matrix takes the same form as Eq. \ref{eix} but with $a,b$ one of
four values, $(1,\ua), (1,\da), (2,\ua), (2, \da)$.  Again 
we have $b(p,q)/c(p,q) = i(g(p) - g(q))$, sufficient for the Yang-Baxter relation to be satisfied, but with
$g(p) = (p - \epsilon_d - U/2)^2/(V^2U)$.

To construct the N-particle eigenfunctions, we again employ a nested Bethe ansatz \cite{original}.  
Crucial to these eigenfunctions being of the Bethe form are both the projectors, $P$, 
{\it and} the correlated hopping term (Eq. \ref{eii}) of ${\cal H}_{2}$.  
This demonstrates that in the case of dot systems with 
orbital degeneracies and a
finite $U$ Coulomb repulsion,
and unlike Hubbard models with orbital degeneracies, it is possible to find a simple
Hamiltonian which is exactly solvable.  
As was demonstrated in Ref. \cite{halchoy}, the ($N>2$)-particle wavefunctions
of any simple finite U orbitally degenerate Hubbard model are not of the Bethe type.

The quantization conditions of an N-electron state carrying momenta $\{ q_i\}^N_{i=1}$ are
of the form
\begin{eqnarray*}\label{exiva}
e^{iq_jL+i\delta_2(q_j)} \!\!&=&\!\! \prod^M_{\alpha = 1}\frac{\lambda^0_j\!-\!\lambda^1_\alpha\!+\!\frac{i}{2}}
{\lambda^0_j\!-\!\lambda^1_\alpha\!-\!\frac{i}{2}};\cr
\prod^{M_{k-1}}_{\lambda^{k-1}_\gamma}
\frac{\lambda^k_\alpha\!-\!\lambda^{k-1}_\gamma\!+\!\frac{i}{2}}{\lambda^k_\alpha\!-\!\lambda^{k-1}_\gamma\!-\!\frac{i}{2}} 
\!\!&=&\!\! \prod^{M_k}_{\beta\neq\alpha}\frac{\lambda^k_\alpha\!-\!\lambda^k_\beta\!+\!i}
{\lambda^k_\alpha\!-\!\lambda^k_\beta\!-\!i}
\prod^{M_{k+1}}_{\delta=1}\frac{\lambda^k_\alpha\!-\!\lambda^{k+1}_\delta\!-\!\frac{i}{2}}
{\lambda^k_\alpha\!-\!\lambda^{k-1}_\delta\!+\!\frac{i}{2}}.
\end{eqnarray*}
Here $\delta_2(q) = -2\tan^{-1}(V^2/2(q-\epsilon_d))$.  The quantum numbers, $\{\lambda^k_\alpha\}$, $k=0,1,2,3$
(with $\lambda^0 = g(q)$) correspond to both spin and orbital degrees of freedom.
In a preliminary analysis of Eqns. \cite{konik2}, we have verified the SU(4) Kondo 
physics expected in a regime where one electron sits on the dots \cite{su(4)}.  Thus
the marginal correlated hopping term in Eqn. (\ref{eii}) does not influence 
the universality class into which the physics falls.  We also note that the integrability of {\bf SD} can be
generalized to N-dots arranged in triangles (N=3), squares (N=4), etc.

\noindent{\bf T=0 Conductance of PD}:
In the remainder of the paper, we focus on extracting the features of the $T=0$ linear response conductance
of the parallel dots ({\bf PD}).  The structure of the linear response conductance, $G$, for the parallel
dots is much richer than that of a single dot containing distinct Kondo effects, novel
applications of the Friedel sum rule, and quantum critical behavior.

To compute the linear response conductance of the dots we closely follow Ref. \cite{konik}.  The
approach is based on the observation \cite{andrei} that the impurity scattering phase, $\delta_{\rm imp}$, 
of an electron is determined by the shift in the electron's momentum due to the presence of the impurity,
i.e $p \rightarrow p + \delta_{\rm imp}/L$.  From the Bethe
ansatz the full momentum of the excitations are readily extracted.  Then isolating the term in
the momentum scaling as the inverse system size, $L^{-1}$, allows the $\delta_{\rm imp}$ to be computed.
In the geometry we have chosen, $G$ is given by $2e^2/h\sin^2(\delta_{\rm imp}/2)$.
For a detailed description in the context of {\bf PD}s see Ref. \cite{konik}.

\begin{figure}[tbh]
\includegraphics[height=2in,width=3.in,angle=0]{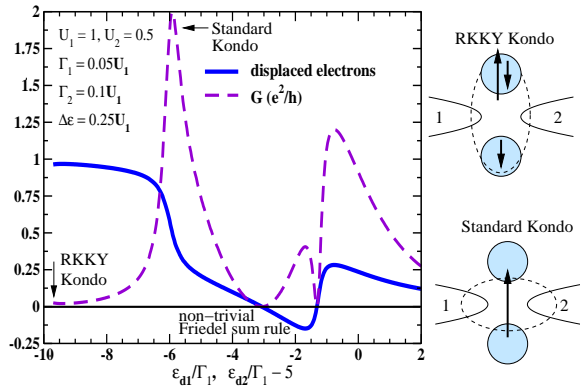}
\caption{The total conductance and number of displaced electrons per spin species vs. $\epsilon_{d1}$,
$\Delta\epsilon$ fixed, for two dots in parallel.}
\end{figure}

We plot an example of the linear response conductance for two dots in parallel in Fig. 2.
We consider the asymmetric case where
$\Delta\epsilon = \epsilon_{d1}- \epsilon_{d2} \gg \Gamma_1,\Gamma_2$.  The conductance
is plotted as a function of $\epsilon_{d1}$ keeping $\Delta\epsilon$ fixed 
from the particle-hole symmetric (p.h.s.) point of the system (i.e. $U_\alpha = -2\epsilon_{d\alpha}$)
where two electrons sit on the dots to a point where both dot levels are well above the Fermi
level and the dots are nearly empty. 
In this plot, the structure of the linear response conductance, in comparison with 
that for a single level dot, is more complex.  This reflects both the presence of interference
\cite{kawa} as well as distinct
types of Kondo physics.

At the p.h.s point, (i.e. $\epsilon_{d1}/\Gamma_1 = -10$ in Fig. 2) the
two electrons residing on the dots together with electrons in the leads form a singlet.  
With no bare direct exchange, singlet formation is mediated solely by virtual hopping processes which
here promote {\it antiferromagnetic} correlations.  
We thus term this formation the RKKY-Kondo effect to mark the role
of electron itinerancy.  This phenomena is distinct from the {\it ferromagnetic} RKKY effect 
arising in fourth order perturbation theory for two closely spaced dots and is thought to
compete with Kondo physics.  In particular, the RKKY-Kondo effect
is non-perturbative in virtual hopping processes.  We find that the RKKY-Kondo effect
is generically present provided $\epsilon_{d1}\neq \epsilon_{d2}$ and so there are unequal
numbers of electrons on each dot (as indicated in the cartoon in Figure 2).  

This antiferro-RKKY Kondo effect is, in a sense, unsurprising.  The Friedel sum rule (FSR) dictates
that the at the p.h.s. point, the scattering phase equal $\pi$ and $G$ vanish.
If we were instead to have a ferro-RKKY effect, and so an underscreened spin 1 impurity,
$G$ would be maximal and so violate the FSR.
Our finding of a Fermi liquid fixed point is supported indirectly by earlier work on the two impurity
Anderson model \cite{jones}.  Although there the focus is upon two electron channels coupled to two impurities, 
the finding is both that the physics is Fermi liquid and that the one to two channel
crossover is smooth.  We have verified singlet formation
both by demonstrating from the Bethe ansatz equations
that the entropy vanishes in the zero temperature limit
and, separately, from a slave boson mean field analysis.

The Abrikosov-Suhl resonance associated with this RKKY-Kondo effect
can be computed along the lines of Ref. \cite{konik}  (up to a multiplicative
constant):
$$
\rho(\epsilon) = \cos(\beta\pi)(T_{RK}^{-1}(\tilde\epsilon^2+1))/(2\tilde\epsilon^2\cos(2\beta\pi)+\tilde\epsilon^4 +1),
$$
where $\tilde\epsilon = \epsilon/T_{RK}$ and $T_{RK}$ is
the RKKY Kondo temperature, 
$T_{RK} \sim \sqrt{U_1\Gamma_1}\exp(-\pi U/8\Gamma_1)$. The parameter $\beta$ is
$~0$ if $\Delta \epsilon \gg \Gamma_{1,2}$ and is
$~1/2$ if $0 < \Delta \epsilon \ll \Gamma_{1,2}$.  As $\Delta\epsilon\rightarrow 0$,
the resonance evolves from a Lorentzian centered at zero energy to a structure with split peaks.
The scale $T_{RK}$ governs 
the leading corrections to the conductance as a function of Zeeman field (H) and
temperature (T).  For $\Delta \epsilon \gg \Gamma_{1,2}$, 
we compute these at the p.h.s. point to have the Fermi-liquid form
(in units of $2e^2/h$):
$G(T/T_{RK}) = \pi^4(T/T_{RK})^2/4$ and $G(H/T_{RK}) = \pi^2(H/T_{RK})^2/4$.

As we move away from the p.h.s. point through increasing the gate voltage,
we begin to empty the dots. With the assumed asymmetry in $\epsilon_{d1}$ and $\epsilon_{d2}$,
we arrive at a point where the dot system has roughly one electron sitting predominantly on
dot 1.  At this value of the gate voltage, we expect ordinary Kondo physics
to be operative.  Defining $T_K$ via its relation
to the static impurity susceptibility, i.e. $\chi_{\rm imp}= (4T_K)^{-1}$,
the corresponding leading
contributions to the conductances (in units of $2e^2/h$) are
$G(T/T_{K})=1-(\pi^4/16)(T/T_{K})^2$ \cite{costi,konik} and $G(H/T_{K})=1-(\pi^2/16)(H/T_{K})^2$ \cite{konik}.  
From a numerical analysis, we know that $T_K$ has a single dot form, i.e.
$T_{K} \sim \sqrt{U\Gamma_1/2}\exp(\pi\epsilon_{d1}(\epsilon_{d1}+U)/2U_1\Gamma_1)$ \cite{original,haldane}.

As $\epsilon_{d1}$ is further increased, we see both a vanishing of
the conductance and an unusual form of the FSR, a mark of the effects of interference.  The vanishing of
the conductance may reflect interference alone: it is present as well in the non-interacting case.   
However the form the FSR takes reflects both interference and interactions.  The FSR relates the scattering
phase to the number of displaced electrons, i.e. $\delta_{e\sigma} = \pi n_{{\rm dis}\sigma}$.
$n_{{\rm dis}\sigma}$ is defined to be
\begin{equation}
n_{{\rm dis}\sigma} \equiv n_{d\sigma} + \int dx \bigg[ \langle c^\dagger_{e\sigma}(x) c_{e\sigma}(x)\rangle - \rho_{{\rm bulk}\sigma}\bigg],
\end{equation} 
and contains contributions from both the occupancy of the dots, $n_{d\sigma}$, and
deviations in the lead electron density from coupling the dots to the leads \cite{langreth}.
As is evident in Figure 2, we have an unusual situation where both contributions to $n_{{\rm dis}\sigma}$,
and not merely $n_{d\sigma}$, are finite:  $n_{d\sigma}$ is always manifestly positive while $n_{{\rm dis}\sigma}$ is negative
over a range of $\epsilon_{d1}$.

One last feature to the linear response conductance we wish to point out 
is the disappearance of the Abrikosov-Suhl resonance
at precisely $\epsilon_{d1}=\epsilon_{d2}$.  The transition is first order
as $T_{RK}$ itself does not vanish as $\epsilon_{d1}\rightarrow \epsilon_{d2}$.  
The origin of this critical point lies in the 
decoupling of one dot degree of freedom (d.o.f) 
if $\epsilon_{d1}=\epsilon_{d2}$ (and only if), as can be seen
via a change of basis $d_{e/o}=(\Gamma_{1/2}d_1\pm\Gamma_{2/1}d_2)/(\Gamma_1^2+\Gamma_2^2)^{1/2}$.
This discontinuous
behavior at $\epsilon_{d1}=\epsilon_{d2}$ however can be transformed into a smooth crossover by weakly coupling
a second channel of electrons to the dot \cite{meden} and so recoupling the odd dot d.o.f.

Apart from the behavior at $\epsilon_{d1}=\epsilon_{d2}$, 
we generally expect the above physics to be robust against small violations of the integrability
constraints (Eqn. 9) for a number of reasons \cite{konik2}: i) the ground state of the dot-lead system is 
already robustly established (unlike when perturbation theory in $V$ is done for a single dot-lead); and
ii) a Schrieffer-Wolfe transformation in the Kondo regime 
is unaffected by (weak) violations of Eqn. (7).


RMK acknowledges support from the US DOE
(DE-AC02-98 CH 10886) together with useful discussions with A. Tsvelik and F. Essler.


\begin{references}

\bibitem{RKKY1} 
H. Jeong et al., Science {\bf 293} 2221 (2001). 

\bibitem{RKKY2} N. J. Craig et al., Science {\bf 304}, 565 (2004). 

\bibitem{scdots1}
J. Petta et al., Science {\bf 309} 2180 (2005).

\bibitem{scdots2} F. Koppens, et al., Science {\bf 309}, 1346 (2005).

\bibitem{scdots3} J.C. Chen et al., Phys. Rev. Lett. {\bf 92}, 176801 (2004).

\bibitem{scdots4} M. Sigrist et al., Phys. Rev. Lett. {\bf 93}, 066802 (2004).

\bibitem{cndots1}
N. Mason et al., Science {\bf 303} 655 (2004). 

\bibitem{cndots2} S. Sapmaz et al., cond-mat/0602424.

\bibitem{gold} D. Goldhaber-Gordon et al.
PRL {\bf 81}, 5225 (1998);

\bibitem{kondo}
S. Cronenwett et al., Science {\bf 281}, 540 (1998).

\bibitem{costi} T. Costi et al., J. Phys.: Cond. Mat. {\bf 6}, 2519 (1994).

\bibitem{konik} R. Konik, H. Saleur, A. Ludwig,
Phys. Rev. Lett. {\bf 87}, 236801 (2001); ibid, Phys. Rev. B {\bf 66}, 125304 (2002).

\bibitem{glazman} M. Vavilov, L. Glazman, Phys. Rev. Lett {\bf 94}, 086805 (2005).

\bibitem{oreg} P. Simon et al., Phys. Rev. Lett. {\bf 94}, 086602 (2005).

\bibitem{lopez} R. L\'opez et al., Phys. Rev. Lett. {\bf 89}, 136802 (2002).

\bibitem{meir} A. Georges, Y. Meir, Phys. Rev. Lett. {\bf 82}, 3508 (1999).


\bibitem{kawa} Y. Tanaka and N. Kawakami, cond-mat/0503341.

\bibitem{su(4)} L. Borda et al., Phys. Rev. Lett. {\bf 90}, 026602 (2003).

\bibitem{tubes} P. Jarillo-Herrero et al., Nature {\bf 434}, 484 (2005); M. Choi et al.,
Phys. Rev. Lett {\bf 95} 067204 (2005).

\bibitem{andrei} N. Andrei, Phys. Lett. {\bf 87A}, 299 (1982).

\bibitem{konik2} R. M. Konik, in preparation.; ibid, cond-mat/0701670.

\bibitem{original} A. Tsvelik and P. Wiegmann, Adv. Phys. {\bf 32}, 453 (1983);
N. Kawakami and A. Okiji, Phys. Lett. A {\bf 86}, 483 (1981).

\bibitem{schlottmann} P. Schlottmann, Phys. Rev. Lett {\bf 50}, 1697 (1983).

\bibitem{halchoy} D. Haldane, T. Choy, Phys. Lett. A {\bf 90}, 83 (1981).

\bibitem{haldane} D. Haldane, Phys. Rev. Lett. {\bf 40}, 416 (1978).

\bibitem{jones} B. A. Jones et al., Phys. Rev. B. {\bf 39}, 3415 (1989).

\bibitem{langreth} D. Langreth, Phys. Rev. {\bf 150}, 516 (1966).

\bibitem{meden} V. Meden, F. Marquardt, Phys. Rev. Lett. {\bf 96}, 146801 (2006).




\end{references}
\end{document}